\documentclass[aps,prl,onecolumn,amsmath,amssymb,superscriptaddress]{revtex4-2}
\usepackage{graphicx}
\usepackage{color}
\definecolor{cerulean}{rgb}{0.0, 0.48, 0.65}
\definecolor{regalia}{rgb}{0.32, 0.18, 0.5}
\definecolor{teal}{rgb}{0.0, 0.13, 0.13}
\usepackage{float}
\usepackage{accents}
\usepackage{hyperref}
\usepackage{bbold}
\usepackage{soul}
\hypersetup{
    colorlinks=true,
    linkcolor=regalia,
    filecolor=regalia,      
    urlcolor=regalia,
    citecolor=regalia
}

\usepackage{color}

\usepackage[normalem]{ulem}
\usepackage{subfigure}
 \usepackage{tikz} 
 \usetikzlibrary{decorations.pathreplacing,calligraphy}

\def\be{\begin{equation}}
\def\ee{\end{equation}}
\def\ba{\begin{eqnarray}}
\def\ea{\end{eqnarray}}
\def\bc{\begin{center}}
\def\ec{\end{center}}

\usepackage[colorinlistoftodos,prependcaption,textsize=footnotesize]{todonotes}

\begin{document}
\title{Computing the graph-changing dynamics of loop quantum gravity}

\author{T. L. M. Guedes}
\email{t.guedes@fz-juelich.de}
\affiliation{Institute for Quantum Information, RWTH Aachen University, D-52056 Aachen, Germany}
\affiliation{Peter Gr{\"u}nberg Institute, Theoretical Nanoelectronics, Forschungszentrum J{\"u}lich, D-52425 J{\"u}lich, Germany}

\author{G. A. Mena Marug{\'a}n}
\email{mena@iem.cfmac.csic.es}
\affiliation{Instituto de Estructura de la Materia, IEM-CSIC, C/ Serrano 121, 28006 Madrid, Spain}

\author{F. Vidotto}
\email{fvidotto@uwo.ca}
\affiliation{Instituto de Estructura de la Materia, IEM-CSIC, C/ Serrano 121, 28006 Madrid, Spain}
\affiliation{Department of Physics and Astronomy, Department of Philosophy, and Rotman Institute, Western University,
N6A 5B7, London, Ontario, Canada.}

\author{M. M{\"u}ller}\email{markus.mueller@fz-juelich.de}
\affiliation{Institute for Quantum Information, RWTH Aachen University, D-52056 Aachen, Germany}
\affiliation{Peter Gr{\"u}nberg Institute, Theoretical Nanoelectronics, Forschungszentrum J{\"u}lich, D-52425 J{\"u}lich, Germany}

\begin{abstract}
In loop quantum gravity (LQG), states of the gravitational field are represented by labeled graphs called spin networks. Their dynamics can be described by a Hamiltonian constraint, { which acts on the spin network states modifying both spins and graphs.} Fixed-graph approximations of the dynamics have been extensively studied, but its full graph-changing action so far remains elusive. The latter, alongside the solutions of its constraint, are arguably the missing features { in canonical LQG to access phenomenology in all its richness}. Here, we discuss a recently developed numerical tool that, for the first time, implements graph-changing dynamics via the Hamiltonian constraint. We explain how it is used to find new solutions to that constraint and to show that some quantum geometric observables behave differently than in the graph-preserving truncation. We also point out that these new numerical methods can find applications in other domains.
\end{abstract}

\maketitle

General Relativity was introduced 110 years ago by Einstein to describe the gravitational interaction in a geometric language that extends the universality of Physics beyond the framework of nonacelerated observers \cite{Einstein}. Despite the success of the theory in explaining a wide variety of astrophysical and gravitational scenarios, and the accumulated observational confirmation of its predictions \cite{observations}, the fact that the principles of General Relativity are based on a classical formulation of the spacetime geometry is in conflict with Quantum Mechanics. Combining these two different visions of nature (classical and quantum) into a single theory that includes gravity is probably the main challenge of General Relativity, which still awaits a { consensual} solution \cite{quantum}.

One of the most solid candidates for achieving such a consistent quantum version of General Relativity is Loop Quantum Gravity (LQG) \cite{Thiemann,ashtekar_review,RovelliVidotto}.   
A distinct feature of this quantum theory is that geometric observables acquire discrete spectra \cite{RovelliSmolin94}. Next year (2026) will mark the 40th anniversary of the complete definition by Ashtekar of new variables \cite{variables}, inspired by the works of Sen \cite{Sen}. These variables allow General Relativity to be reformulated as a theory of connections and to build the mathematical structure on which LQG rests. However, while this mathematical structure is quite well defined, computations in this theory present an enormous challenge. In particular, the dynamics are encoded as part of the symmetries and are therefore governed by a (quantum) constraint \cite{Thiemann,ashtekar_review}. This, called the quantum Hamiltonian constraint, has proven to be extremely complex, and its implementation and resolution is the main obstacle to completing the quantization program of LQG.  Recently, several numerical computations~\cite{numerics, Steinhaus, SL2Foam, Muxin, Bianca, Sherif} have been developed  within the covariant formulation of the theory~\cite{RovelliVidotto}, but in general numerical tools remain scarce, particularly in the canonical approach~\cite{Hanno, Ilkka_gc}, where the four-dimensional spacetime manifold is foliated, adopting a 3+1 description \cite{Thiemann}.

More concretely, the dynamics of canonical LQG is defined by a single operator, the aforementioned Hamiltonian constraint \cite{RovelliSmolin93}. It acts on a state space for the spacetime geometry that admits a basis formed by spin networks \cite{Penrose71,RovelliSmolin95}. These are graphs with links carrying spin labels that (normally) form singlets at the nodes~\cite{temperley, Brink, Messiah, Ilkka}. They find application in several areas of physics and quantum computing~\cite{Aquilanti}. The Hamiltonian constraint changes the graph of a spin network and its spin assignments in multiple ways, yielding a superposition of spin networks with different graphs.  This complication is amplified by the presence of the volume operator in the Hamiltonian. Calculating matrix elements of the volume requires diagonalizing matrices built on subspaces of spin networks with identical graphs~\cite{Lewandowski_qvolume1,Lewandowski_qvolume2,Thiemann_vol,Giesel_qvolume1,Giesel_qvolume2,Ma_vol,DePietri}, which demands developing new numerical approaches. The resulting dynamics, as well as the solutions to its constraint, have therefore remained inaccessible. Even its effect on the volume -- a key geometric observable -- has yet to be characterized, preventing canonical LQG from reaching the physically correct quantum domain.

\begin{figure}[t]
    \centering
    \includegraphics[width=0.75\linewidth]{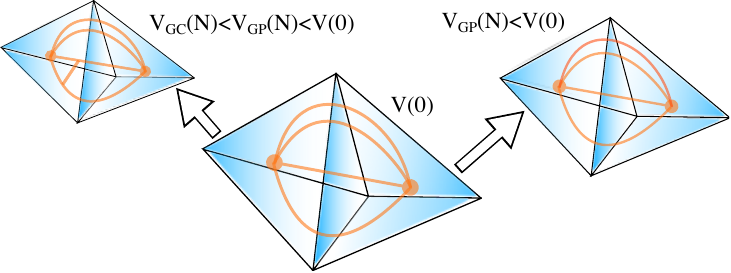}
 \caption{The dipole model as an example of a spin network. The corresponding spin network consists of two $4$-valent nodes sharing each of their four links. Its dual is formed by two tetrahedra with faces that are pairwise glued (in four dimensions). These tetrahedra represent quanta of volume in a discretized geometry. The unitary formed by exponentiating the Hamiltonian with the lapse $N$ as a perturbative parameter transforms the dipole model differently when graph-changing (left) or graph-preserving (right) dynamics are considered. All figures in this work are based on figures of Ref. \cite{Companion}.}
  \label{fig1}
\end{figure}

In this work, we discuss and summarize the first numerical approach implementing the Hamiltonian constraint on 3- and 4-valent spin networks,  dual graphs to triangulations of respectively two- and three-dimensional hypersurfaces (``spacetime cuts"), without recurring to truncations to fixed graphs as is common in LQG (cf. Fig.~\ref{fig1}). The valence of the spin network is the number of links joining at (each of) its vertices. This approach has been recently introduced in Ref.~\cite{Companion}. Given the length and complexity of the calculations presented in that reference, we provide here a concise and much more pedagogical presentation of the main results of that work, reviewing the concepts that are necessary to understand them, and discussing the implications and potential applications of this important contribution towards the completion of LQG.  The new numerical approach allows for recursive application of the Hamiltonian constraint on spin networks, yielding perturbative expansions of constraint-generated operators. A key feature is a bijection between  spin networks and functions of lists, on which the action of the Hamiltonian is implemented as a functional. We explain how the applicability of this approach may reach beyond LQG. For the 3-valent case, akin to Ref.~\cite{Ma}, the derivations update those in Refs.~\cite{borissov, gaul}. Furthermore, we discuss the first complete calculations of the action of the Hamiltonian on spin networks of valence four.  

The work focus on the numerical (reference-frame free) study of graph-changing canonical LQG, computing volume expectation values of two perturbatively transformed 4-valent spin networks as an illustrative application with important physical consequences. The results are then compared with their counterparts originating from a graph-preserving Hamiltonian, presenting concrete indication that the latter fails to capture the proper dynamics in LQG. These results provide the missing reference point to devise and test approximations to the graph-changing dynamics, and should enable to perform certain calculations without approximations. Lastly, we discuss the construction of an infinite family of solutions to the Hamiltonian constraint. Until now, no solutions were known without additional assumptions~\cite{Lewandowski_state, Pullin_state, Madhavan}. Performing graph-changing computations opens a range of possibilities, including checks about how graph-changing formulations affect semi-classical predictions~\cite{bounce, Cong}.

Using canonical variables consisting of densitized triads and $su(2)$-connections~\cite{variables, Lee_knots, Barbero}, the gravitational action can be rewritten (up to a Legendre term) as a sum over smeared constraints associated with three kinds of symmetries: gauge invariance, diffeomorphisms and time reparametrization~\cite{thiemann, QSD, QSD2, ashtekar_review}. This system can be quantized ``\`a la Dirac", promoting the constraints to operators~\cite{ashtekar_review, Dirac2}.  The gauge constraints are satisfied by considering only spin networks in which all nodes correspond to spin singlets  (also called intertwiners). For spin networks embedded in manifolds, diffeomorphisms can be understood as (invertible analytic) deformations of graphs. To satisfy the diffeomorphism constraints, one considers equivalence classes of (dual) spin networks~\cite{rovelli2004book, ashtekar_review}: all graphs related by these deformations should be superposed. The last constraint, referred to as Hamiltonian, dictates the dynamics. Here, we will consider the so-called Euclidean version of this constraint, as a first but fundamental step towards the determination of its Lorentzian counterpart. The solutions to all those constraints provide the physical states. Finding these states without any approximations or additional assumptions has remained an open problem in LQG.  

\begin{figure}[t]
    \centering
    \includegraphics[width=0.955\linewidth]{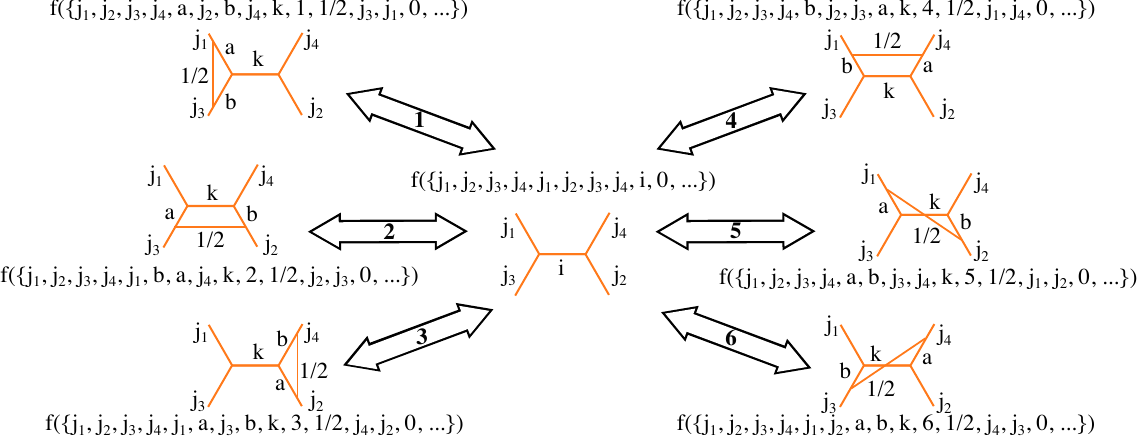}
 \caption{ A $4$-valent spin network node (center), with its assigned ghost function above, is mapped under the action of the Hamiltonian into six modified structures which contain inner loops. Ghost functions containing the lists encoding such spin networks are given. Double arrows emphasize the reversible character of the Hamiltonian, and the numbers within them highlight the location of the added loop. All figures in this work are based on figures of Ref. \cite{Companion}.}
  \label{fig2}
\end{figure}

According to our comments above, in pure gravity we consider the Hamiltonian~\cite{borissov, QSD}: 
\begin{equation}
    \hat{C}_s = \lim_{\boxtimes \to 0} \sum_{\boxtimes} \frac{iN\epsilon_{ijk}}{3 l^2_0} \text{tr}  \left\{ \hat{h}[\alpha_{ji}]-\hat{h}[\alpha_{ij}], \hat{h}[p_k] \hat{V} \hat{h}^{-1}[p_k] \right\} .
    \label{scalar_constr}
\end{equation} 
Braces denote the anticommutator, $\epsilon_{ijk}$ is the totally anti-symmetric symbol, $l_0$ is the Planck length and $\text{tr}$ is the trace. The operators in Eq.~\eqref{scalar_constr} are the volume $\hat{V}$ and the holonomies $\hat{h}[p]$, parallel-transport unitaries over the path $p$. We represent by $\boxtimes$ a partition of the considered manifold in three-dimensions into tetrahedra, with sizes approaching zero, $\boxtimes \to 0$, while their number diverges. As a result, only tetrahedra with vertices at spin network nodes and spanned by $\alpha_{ij}$ and $p_k$ contribute, and no tetrahedron contains more than one node~\cite{thiemann, QSD}. The prefactor $N=N_\boxtimes$ is given by the evaluation at each node of the lapse of the spacetime foliation (or triangulation thereof). This lapse modulates the action of the constraint in each tetrahedron. Since we can study each node separately, only one $N_\boxtimes=N$ contributes in our following considerations. The loops $\alpha_{ij}$ and $\alpha_{ji}=\alpha^{-1}_{ij}$ of opposite orientations are formed by segments tangent to two links (labeled $i$ and $j$), connected by an additional link that produces a graph-changing effect (see Fig.~\ref{fig2}). The path $p_k$ corresponds to a segment tangent to another link from the node, to which the label $k$ is assigned. The holonomies in Eq.~\eqref{scalar_constr} couple additional spins to the links on which the Hamiltonian acts. Lastly, the volume operator gives the number of quanta of volume, which depends on the spins of the links connected to nodes of valence higher or equal to four~\cite{Lewandowski_qvolume2}. This provides an interpretation of spin networks as duals to triangulations, associating a link to each face, and a node to each tetrahedron (see Fig.~\ref{fig1}). The volume also depends on the relative arrangement of links at each node~\cite{Lewandowski_qvolume2}, therefore we restrict our analysis to  spin networks with links oriented along the faces of a tetrahedron (so that each triplet of links is linearly independent). The effect of the volume or the constraint on these spin networks  should not be influenced by diffeomorphisms (or averaging over them) ~\cite{QSD}. Likewise, the matrix elements of any geometric operator, such as lengths or dihedral angles, evaluated on diffeomorphism averaged (dual) spin networks, should be diffeomorphism-invariant quantities that can be calculated by choosing any representative from a family of diffeomorphism-equivalent spin networks, since the extent of these quantities is determined by the link spins.

The constraint $\hat{C}_s$ has the same action as the symmetric Hamiltonian proposed by Thiemann~\cite{QSD2}: when acting  at a node, whenever the Hamiltonian  introduces a loop in the location where the deepest inner loop is, these loops get coupled. As a result, the spin of a loop link can be decreased, leading to link removal when its spin reaches zero. In Ref.~\cite{QSD2}, it is explained that this is a subtle caveat, crucial to obtain a formally symmetric operator on spin networks given by unitary holonomies and the volume, and guarantees that the domain of this Hamiltonian operator is densely defined.  

Transforming spin networks according to the symmetric Hamiltonian demands storing information about superpositions of different graphs with spins assigned to their links, while allowing them to assume increasingly more complex structures resulting from loops added by the Hamiltonian closer and closer to the central nodes. The proposed solution is to store spin and location information as ordered lists, each in one-to-one relation with an spin network~\cite{Companion}. This allows for the construction of a vector space of abstract functions for which the arguments are these lists. These are called ghost functions because they are never assigned a functional form. If $s_i=\{s_{i,1}, s_{i,2}, \ldots \}$ denotes lists encoding spin networks, the functions $f(s_i)$ are endowed with the inner product $I[ f(s_i), f(s_j) ] = \delta_{ij} $. The Hamiltonian is then coded as a linear functional that acts on ghost functions by reading and changing their arguments: $C_s [f(s_i)]=\sum_j c_j(s_i) f(s_j)$ for coefficients $c_j$ originating from the action of Eq.~\eqref{scalar_constr} on 3-{valent} or 4-valent spin networks and derived in Ref.~\cite{Companion}. Linearity implies $C_s [\sum_i c_i f(s_i)]=\sum_i c_i C_s[f(s_i)]$, thus the functional can be used recursively. 

\begin{figure}[t]
    \centering
    \includegraphics[width=0.95\linewidth]{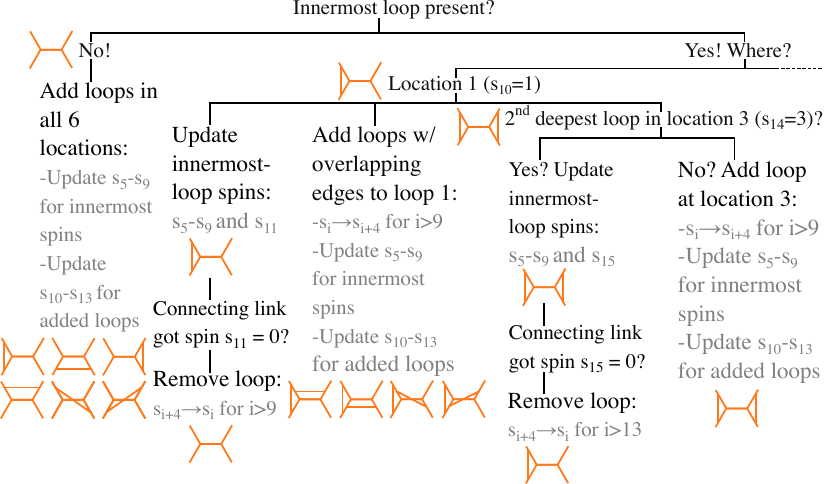}
 \caption{Pseudocode exemplifying the implementation of the Hamiltonian. The code first checks whether an inner loop is present. If absent, it generates a linear combination of graphs with inner loops introduced in all six locations, with spin $1/2$ on the newly created link. If present, a series of steps is followed for each possible location (the case for location 1 is explicitly  shown, while for other locations similar rules are implied by the dashed-line continuation of the diagram to the right). Namely, coupling a new loop at the innermost-loop location merely alters spins without affecting the graphs. If the spin of the connecting link becomes zero, it is removed, and the inner-loop data in the corresponding list is shifted to the left by four entries. Also, inner loops are introduced (deeper) in all other positions, but if a loop was inserted at position $3$ right before adding one at position $1$ (these loops share no links), it either removes its extra link or changes spins. The diagram displays examples of the simplest  spin networks (orange graphs) for which the rules apply. All figures in this work are based on figures of Ref. \cite{Companion}.}
  \label{fig3}
\end{figure}

We focus our discussion on the study of 4-valent spin networks with an inner virtual link, four external legs, and an arbitrary number of inner loops. By connecting links belonging to each possible pair of directions, the inner loops can be arranged in six different manners (cf. Fig.~\ref{fig2}). We label such inner-loop locations from 1 to 6, connecting links along the respective pairs $\{p_1,p_3\}$, $\{p_2,p_3\}$, $\{p_2,p_4\}$, $\{p_1,p_4\}$, $\{p_1,p_2\}$, and $\{p_3,p_4\}$. In the central spin network of Fig.~\ref{fig2}, $j_i$ is the link spin along direction $p_i$. The manner in which Eq.~\eqref{scalar_constr} can attach new loops is affected by the presence of  inner loops in certain locations. If a loop is present, e.g., in location 1 (placed between directions $p_1$ and $p_3$), the constraint attaches a new loop in the same location by coupling its holonomies with the already existing loop links, without changing the graph, but altering the spins of these links (with the exception of the case in which the spin of the connecting link becomes zero, changing the graph as a consequence of the link removal). The Hamiltonian also forms inner loops in all other locations, but the presence of a loop in location 1 implies that loops in locations 2, 4, 5, and 6 (sharing a common link with loop 1) need be introduced further inwards, relative to the loop in location 1. Meanwhile, a loop introduced in location 3 is unaffected by that loop and can be located at similar depth. Consequently, successive application of Eq.~\eqref{scalar_constr} generates graphs with progressively deeper inner loops at depths dependent on loop positions. Figure~\ref{fig3} shows a pseudocode exemplifying the addition/removal of loops.

The (ghost-function) lists have the spins of the outermost links as their first four entries. The next four entries are the innermost spins adjacent to the central (virtual) link, i.e., along directions $p_1, p_2, p_3$, and $p_4$. The ninth entry is the spin of this central virtual link. In case the spin network has no inner loops, the remaining entries have vanishing values (see Fig.~\ref{fig2}). Otherwise, the innermost-loop data are encoded in the next four entries, and each following inner loop, in decreasing  depth order, is described by an additional quartet of entries. In each quartet, the first two entries store respectively the loop location and the spin of its connecting link, while the other two store the spins adjacent to (but not contained in) the loop along the directions it connects. When the constraint creates a new loop, it shifts all entries from the tenth position onwards four places to the right, so that inner-loop entries are moved down in depth order to allow for inclusion of data of the new loop. The new spins adjacent to the central nodes are encoded in entries fifth-eighth, and the new central spin in the ninth entry. Information about the added innermost loop occupies entries tenth-thirteenth. Although spin networks and their encoding lists become increasingly complex, the constraint acts only upon the two deepest inner loops of a 4-valent spin network. Since the information about these two loops is stored in entries fifth to seventeenth, the coefficients $ c_j(s_i)$ in $C_s [f(s_i)]=\sum_j c_j(s_i) f(s_j)$ depend solely on those entries, avoiding the search for data scattered among large lists.

The constraint is implemented as a map between non-normalized spin networks. So, normalization is required after its recursive action. The ``normalizer'' functional linearly implements this according to $f(s_i)\to [d_{j_1}d_{j_2}d_{j_3}d_{j_4}\prod_k d^{-1}_k ]f(s_i)$~\cite{Companion}, where $d_{j}=2j+1$. Here, $j_i$ represent the spins of the outermost legs, and $k$ runs over the spins of all  spin network links, including the outermost ones. To achieve this, the normalizer reads in each ghost-function argument the first six entries and the $j$th, $(j-1)$th, and $(j-2)$th entries for $j = 4n+9$ ($n\in \mathbb{N}$). 

One possible way of understanding the dynamics generated by the Hamiltonian is by investigating the change it causes on geometrical operators for some fiducial spin networks. For this purpose, the code constructed in Ref.~\cite{Companion} also implements a volume functional. It only sees the spins adjacent to the central  spin network link. These spins fix the size of the matrix generated by the volume operator. The volume maps a 4-valent spin network with a central spin $i$ 
into a linear combination of 4-valent spin networks with all possible spins for the central links that are allowed by triangularity, so that the size of the matrix it generates runs from $\min \{ |j'_1 - j'_3|, |j'_2-j'_4|\}$ to $\max \{ j'_1 + j'_3, j'_2+j'_4\}$ (for innermost spins $j'_1, j'_2, j'_3$, and $j'_4$), and the matrix indices correspond to the input and output values of the spin of the central link. The volume is derived from an intermediate operator acting on the spin networks. Its matrix needs to be diagonalized, so that the absolute value and square root of its entries can be taken before the application of  the inverse of the diagonalizing transformation, giving volume matrix elements in a basis of 4-valent spin network states~\cite{Companion}.   

In general, solutions to the constraint \eqref{scalar_constr}, which provide physical states in LQG in the absence of matter, remained unknown until now. Some solutions were found (a) in the presence of a semi-classical massive scalar field~\cite{Lewandowski_state}, (b) using the Temperley-Lieb algebra~\cite{temperley,Pullin_state} and (c) using an incomplete Hamiltonian~\cite{Alesci}, yet none holds in the case we study. The existence of certain classes of diffeomorphism-invariant solutions was also proven, but none was explicitly constructed~\cite{ Madhavan}. Using the coded Hamiltonian, it has been possible to perform a search for states annihilated by it. The protocol runs over semi-natural spins within $[0,7/2]$ on each link of an spin network without inner loops~\cite{Companion}. Only gauge-invariant states are allowed. Within the investigated spin range, we have found a solution only for vanishing $j_1,j_2,j_3,j_4$, and $i$, suggesting that spin networks with zero innermost spins connected to the intertwiners provide solutions.  On the other hand, when the Hamiltonian acts on a linear combination of spin networks that cannot be related by inner-loop couplings, it produces a new linear combination of spin networks orthogonal to the initial state $|s_0\rangle$. Denoting $|s_i\rangle $ the state generated by $i$ loop insertions on $|s_0\rangle$, with $\langle s_i | s_j \rangle = \delta_{ij}$ and $\hat{C}_s |s_0\rangle =c^*_{1} |s_{1}\rangle$, we have
\begin{equation}
\hat{C}_s |s_i\rangle = c_i |s_{i-1}\rangle + c^*_{i+1} |s_{i+1}\rangle =\langle s_{i-1} |\hat{C}_s |s_i\rangle |s_{i-1}\rangle +\langle s_{i+1}|\hat{C}_s |s_i\rangle|s_{i+1}\rangle.
\end{equation}
Therefore, from $|s_0\rangle$,  the following solution is found~\cite{Companion}:
\begin{equation}
    | E_0\rangle = |s_0\rangle  + \sum_{i \geq 1 }(-1)^{i} \bigg[\prod_{j=1}^{i}\frac{\langle s_{2j-1}|\hat{C}_s |s_{2j-2}\rangle }{\langle s_{2j-1}|\hat{C}_s |s_{2j}\rangle }\bigg]  | s_{2j}\rangle \, .
\end{equation}
Note that convergence of this series is not necessary. In fact, a suitable habitat for the infinite family of solutions of the form (2) (corresponding to the infinite possibilities for the choice of $|s_0\rangle$) is the algebraic dual of the linear span of  spin network states, a dual which contains states of infinite norm. If all relevant solutions belong to this dual space, it would suffice to endow it with a convenient (alternative) inner product (and average over diffeomorphims) to build a true Hilbert space of physical states.

We turn now to the discussion of the validity of the graph-preserving approximation commonly used in the literature. By perturbatively transforming spin networks, we estimate how the expectation value of the volume transforms when confronting graph-changing with graph-preserving dynamics. We consider $N$ as our perturbation parameter, and expand (volume matrix elements generated with states evolved by) the unitary $\hat{U}=\exp [-i\hat{C}_s(N)]$~\cite{Reisenberger} up to third and fourth order for graph-changing and graph-preserving scenarios, respectively, disregarding contributions of higher-order terms. It is worth remarking that, since $\hat{U}$ is a unitary, its action preserves the norm of our states, which therefore remains the same at any order in our truncations. On the other hand, notice that in our calculations odd-order terms do not contribute to expectation values of observables. Indeed, when transforming states, any spin network graph can only be recovered after an even number of applications of the graph-changing constraint. Furthermore, the volume operator does not affect the graphs of spin networks it acts on. As a result, in the graph-changing scenario $\langle  \hat{C}^n_s \hat{V}^l \hat{C}^m_s\rangle=0 $ for $n+m$ odd and any $l\in \mathbb{N}$, including zero. For graph-preserving dynamics, however, an odd number of applications of the constraint can recover the original spin network, depending on its connectivity~\cite{Companion}. We consider a ladder-type spin network (or a chain-like spin network, if one accounts for the spatial arrangement of the links) with intertwiners that are connected by their upper or lower pairs of legs and loops which are coupled
solely above and below the fiducial intertwiner, disregarding large loops that could be coupled from the sides (which can favor departures from the graph-changing results).

\begin{figure}
    \centering\includegraphics[width=0.65\textwidth]{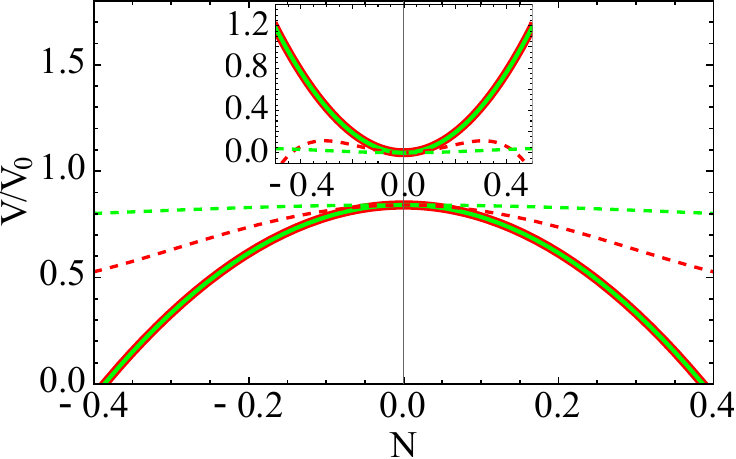}
	\caption{Lapse dependence of the dimensionless volume expectation value. The curves are shown for two spin networks with $j_1 = j_2 = j_3 = j_4 =1/2$, $\varepsilon = 0$, and $i=0$ (red curves) or $i=1$ (green curves). We compare graph-changing (solid) and graph-preserving (dashed) Hamiltonians, for which expectation values are calculated up to third and fourth order in $N$, respectively. Inset: curves for the volume variance. All figures in this work are based on figures of Ref. \cite{Companion}.}
	\label{plot1}
\end{figure}

Figure~\ref{plot1} shows the expectation values of the volume as a function of the lapse $N$ for two fiducial spin networks of the form displayed in the center of Fig.~\ref{fig2}. The chosen spin networks have  $j_1=j_2=j_3=j_4=1/2$ and $i=0$ or $i=1$ (red and green curves, respectively). Unexpectedly, the curves coincide for the two spin networks transformed under graph-changing dynamics. The volumes surprisingly decrease with increasing $|N|$ for $|N|\lesssim 1/2$. For the graph-preserving case, the volume increases when $|N|\gtrsim 1/2$ due to fourth order contributions (although the validity of the results for such values of $|N|$ is questionable in our perturbative approach), and a similar trend is expected for graph-changing dynamics. This represents the first quantitative evidence that graph-preserving approximations lead to departures from the graph-changing dynamics and fail to capture certain symmetries of the system, such as volume degeneracy. The fact that the change appears already at the variation rate around the zero value of the lapse supports the validity of the result, independently of the range of applicability of our perturbative approach. Furthermore, the fact that $\langle  \hat{V}\hat{C}^m_{\tilde{s}}\rangle \neq 0$ can happen in the graph-preserving case for $m$ odd, leading to asymmetries in the volume dependence on the lapse $N$ (and consequently also on the proper time $T=\int \mathrm{d}t N(t)$~\cite{Reisenberger}), clearly shows the severe effects of the graph-preserving approximation. 

Although the spin networks considered in the above calculations are not solutions to the Hamiltonian constraint in vacuo, they can describe the gravitational part of physical states for certain types of matter content, such as a suitable scalar field or nonrotational dust, that may serve as a clock~\cite{Ilkka_gc, Kristina}. In fact, the results translate directly into the case of Gaussian dust~\cite{Kristina}. Consequently, the different volume profiles in the graph-changing and graph-preserving approaches can greatly influence the cosmological phenomenology~\cite{ashtekar_review,LQC}, leading to different expansion rates and inflationary regimes, or in black hole evolution, modifying the black-to-white hole transition time~\cite{WH1,WH2}. It is also worth commenting that, even though only the Euclidean Hamiltonian has been implemented in our analysis, in flat cosmological scenarios this Hamiltonian becomes proportional to the full Lorentzian one~\cite{ABL}. Therefore, one expects the analyzed evolution to capture genuine dynamical features of this type of systems. 

Summarizing, the work that we have reviewed and discussed here is the first solid proposal to introduce numerical tools capable to solve problems involving large superpositions of (changing) graphs. The coded Hamiltonian introduced in Ref.~\cite{Companion} provides quantitative data for the volume, identifying some possible symmetries of the graph-changing dynamics and  showing that it significantly departs from the graph-preserving one. The new numerical avenues open in LQG also allow for the unique finding of new (infinite) families of solutions to the graph-changing Hamiltonian in the absence of matter. The numerics follow the first complete action of the Hamiltonian on 4-valent spin networks (see details in Ref.~\cite{Companion}). These advances enable a new generation of LQG calculations in which approximations are either avoided or better controlled, making it possible to analyze observable properties of quantum geometries, which could eventually be measurable in future experiments \cite{stars,cmb} or quantum simulations.

Finally, let us comment that we expect this computational approach to be useful also beyond LQG. A similar use of ghost functions can be helpful for processes with pair production in cosmology~\cite{Mukha} (if one introduces a cut-off in the excitations), even on classical backgrounds, since in these systems the Hamiltonian dynamics increases at each step the (finite) number of degrees of freedom that must be considered. One such system in which the background can be treated exactly in loop quantum cosmology is the Gowdy model~\cite{Gowdy} (with a cut-off). Our methods can be further extended to treat nonlinear systems of particles confined to lattices (or graphs), in which the number, type and location of the particles is changed by the Hamiltonian (e.g., lattice gauge theories~\cite{Kogut}). More generally, the considered methods can find applications to problems with infinite-dimensional Hilbert spaces spanned by an unstable basis under evolution, but for which the Hamiltonian relates only a finite number of elements per step. Concretely, some amendments to the code should allow for studies of Levin-Wen-type Hamiltonians, which involve graph-changing operations on spin networks, prior to imposing constraints enforcing topological equivalence (akin to the normalizer)~\cite{Levin-Wen, koenig, stringnet_simulation, Fibonacci}. Additionally, computations in many-body systems for states with energy constraints could also be facilitated by encoding the locations and excitation levels in ghost functions~\cite{Igor}. Lastly, we emphasize that, although the presented code is not optimally scalable or efficient when handling spin networks with high spins (cf. Section X of Ref.~\cite{Companion}), scalability and efficiency are an issue for almost any initial numerical breakthrough in every field of physics. Nonetheless, a computationally optimized $C++$ version of the code being currently developed to make use of several compute nodes shows a preliminary speed-up of at least four orders of magnitude for the presented calculations, highlighting how much potential for improvement and development the presented approach still has.

\begin{acknowledgments}
\section{Acknowledgements}
The authors are grateful to Ilkka M{\"a}kinen, Jorge Pullin, Carlo Rovelli and Etera Livine for discussions and suggestions.

FV's research has been partially supported by the Canada Research Chairs Program and the Natural Science and Engineering Council of Canada (NSERC) through the Discovery Grant ``Loop Quantum Gravity: from Computation to Phenomenology''.  She has also been supported by the ID\# 62312 grant from the John Templeton Foundation, as part of the project ``The Quantum Information Structure of Spacetime'' (QISS), https://www.templeton.org/grant/the-quantum-information-structure-ofspacetime-qiss-second-phase. TLMG and MM acknowledge support by the ERC Starting Grant QNets through Grant No. 804247, and the EU’s Horizon Europe research and innovation program under Grant Agreement No. 101114305 (``MILLENION-SGA1” EU Project) and under Grant Agreement No. 101046968 (BRISQ). MM furthermore acknowledges support by the Deutsche Forschungsgemeinschaft (DFG, German Research Foundation) under Germany’s Excellence Strategy ``Cluster of Excellence Matter and Light for Quantum Computing (ML4Q) EXC 2004/1'' 390534769. This research is part of the Munich Quantum Valley (K-8), supported by the Bavarian state government with funds from the Hightech Agenda Bayern Plus. GAMM acknowledges support by MCIU/AEI/10.13019/501100011033 and FSE+ under the Grant No. PID2023-149018NB-C41.

\end{acknowledgments}

\end{document}